\pdfoutput=1
\documentclass[10pt,journal,compsoc]{IEEEtran}
\IEEEoverridecommandlockouts

\ifCLASSOPTIONcompsoc
  \usepackage[nocompress]{cite}
\else
  \usepackage{cite}
\fi

\ifCLASSINFOpdf
  \usepackage[pdftex]{graphicx}
\else
\fi

\usepackage[inline]{enumitem}

\usepackage[]{xcolor}
\usepackage{multicol}
\usepackage{hyperref}
\usepackage[nobreak]{mdframed}
\usepackage{balance}
\usepackage{amsmath}
\usepackage[normalem]{ulem}
\usepackage{booktabs}
\newcommand{\TODO}[1]{\textcolor{red}{#1}\GenericWarning{}{LaTeX Warning: TODO: #1}}\newcommand\todo\TODO

\newcommand{\revision}[1]{\textcolor{black}{#1}}

\usepackage[para,online,flushleft]{threeparttable}
\usepackage{multirow}
\usepackage{makecell}
\usepackage{listings}
\usepackage{xspace}
\usepackage[inline]{enumitem}

\newcommand{\sorald}{\textsc{Sorald}\xspace}
\newcommand{\soraldbot}{\textsc{SoraldBot}\xspace}
\newcommand{\spoon}{\textsc{Spoon}\xspace}
\newcommand{\spongebug}{\textsc{SpongeBugs}\xspace}
\newcommand{\sonar}{\textsc{SonarJava}\xspace}
\newcommand{\sonarqube}{\textsc{SonarQube}\xspace}

\newcommand{\sonarsource}{\textsc{SonarSource}\xspace}

\newcommand{\github}{\textsc{Github}\xspace}

\DeclareUrlCommand\ULurl{%
  \renewcommand\UrlLeft{\uline\bgroup}%
  \renewcommand\UrlRight{\egroup}}

\newcommand{\target}{target\xspace}
\newcommand{\toprepos}{\textsc{TopRepos}\xspace}

\newcommand{\topreposcnt}{161\xspace}

\newcommand{\GHilight}{\makebox[0pt][l]{\color{green}\rule[-4pt]{1\linewidth}{10pt}}}
\newcommand{\RHilight}{\makebox[0pt][l]{\color{pink}\rule[-4pt]{1\linewidth}{10pt}}}

\definecolor{dkgreen}{rgb}{0,0.6,0}
\definecolor{gray}{rgb}{0.5,0.5,0.5}
\definecolor{mauve}{rgb}{0.58,0,0.82}
\definecolor{gray}{rgb}{0.4,0.4,0.4}
\definecolor{darkblue}{rgb}{0.0,0.0,0.6}
\definecolor{lightblue}{rgb}{0.0,0.0,0.9}
\definecolor{cyan}{rgb}{0.0,0.6,0.6}
\definecolor{darkred}{rgb}{0.6,0.0,0.0}

\definecolor{javared}{rgb}{0.6,0,0} 
\definecolor{javagreen}{rgb}{0.25,0.5,0.35} 
\definecolor{javapurple}{rgb}{0.5,0,0.35} 
\definecolor{javadocblue}{rgb}{0.25,0.35,0.75} 

\lstdefinestyle{diff}{
    escapechar=\%
}

\usepackage{ifthen}
\usepackage{cleveref}

\crefname{customListing}{Listing}{Listings}

\begin{document}

\title{Sorald: Automatic Patch Suggestions for SonarQube Static Analysis Violations}

\author{
Khashayar Etemadi,
Nicolas Harrand,
Simon Larsén,
Haris Adzemovic,
Henry Luong Phu,
Ashutosh Verma,
Fernanda Madeiral,
Douglas Wikström,
Martin Monperrus%

\IEEEcompsocitemizethanks{
\IEEEcompsocthanksitem K. Etemadi, N. Harrand, S. Larsén, H. Adzemovic, H. L. Phu, D. Wikstrom, F. Madeiral and M. Monperrus are with the KTH Royal Institute of Technology, Stockholm, Sweden\protect\\
Email: \{khaes, harrand, slarse, harisa, tailp, dog\}@kth.se,\protect\\ fer.madeiral@gmail.com, martin.monperrus@csc.kth.se
\IEEEcompsocthanksitem A. Verma was with IIT Bombay, Mumbai, India\protect\\
Email: ashutosh1598@iitb.ac.in
}
}

\IEEEtitleabstractindextext{
\begin{abstract}
Previous work has shown that early resolution of issues detected by static code analyzers can prevent major costs later on. However, developers often ignore such issues for two main reasons. First, many issues should be interpreted to determine if they correspond to actual flaws in the program. Second, static analyzers often do not present the issues in a way that is actionable.
To address these problems, we present \sorald: a novel system that devise metaprogramming templates to transform the abstract syntax trees of programs and suggest fixes for static analysis warnings. Thus, the burden on the developer is reduced from interpreting and fixing static issues, to inspecting and approving full fledged solutions. \sorald fixes violations of 10 rules from \sonarqube, one of the most widely used static analyzers for Java. 
\revision{We evaluate \sorald on a dataset of 161 popular repositories on \github. Our analysis shows the effectiveness of \sorald as it fixes 65\% (852/1,307) of the violations that meets the repair preconditions. Overall, our experiments show it is possible to automatically fix notable violations of the static analysis rules produced by the state-of-the-art static analyzer \sonarqube.}
\end{abstract}

}

\maketitle

\IEEEdisplaynontitleabstractindextext

\IEEEpeerreviewmaketitle

\section{Introduction}
\label{sec:intro}

\IEEEPARstart{S}{tatic} analysis tools are widely used by developers to maintain and improve code quality~\cite{vassallo2020developers,beller2016analyzing,habib2018many}. A static analyzer inspects code without requiring code execution, and can therefore run faster and more predictably than dynamic analyzers, that do require the code to execute~\cite{artho2005combined}. The analysis is often based on a set of rules that can pertain to anything from style conventions~\cite{loriot2019styler} to detecting bugs~\cite{beller2016analyzing} or highlighting common security vulnerabilities\footnote{\url{https://rules.sonarsource.com/java/type/Vulnerability}}. If the source code under analysis breaks a rule, the static analyzer reports a \emph{rule violation}, highlighting which rule is broken and the part of the source code that is responsible. A developer can then take action to fix the violation in order to keep the source code quality from deteriorating. This is an important activity as source code defects and bugs can have dire consequences for program reliability, to the point of causing fatalities~\cite{zhivich2009real} or huge financial costs \cite{zhivich2009real}.

A fundamental problem with the use of static analyzers is that developers tend to ignore rule violations~\cite{liu2018mining}. There are two predominant reasons for this. First, static analyzers have a tendency to overwhelm developers with violations that may not actually pose significant problems, may not be relevant in the context, or may simply be incorrectly identified~\cite{johnson2013don,ernst2015measure}. Second, even assuming that addressing a given violation is worthwhile, interpreting violation messages can be difficult~\cite{johnson2013don,ernst2015measure}.

As developers need to both interpret rule violation messages and implement fixes for them, the burden that static analyzers impose can be high. In order to reduce it, we introduce \sorald\footnote{\sorald means \textbf{so}nar \textbf{r}epair mångf\textbf{ald} (Swedish word for diversity). This reflects both the tool's purpose, and that it is developed by a culturally diverse team with roots all over the world.}, an \emph{automatic program repair} tool for Java programs that fixes rule violations raised by \sonarqube\footnote{\url{https://www.sonarqube.org}}, one of the most popular static analyzers~\cite{saarimaki2019accuracy}. \sorald leverages the underlying \sonar\footnote{\url{https://github.com/SonarSource/sonar-java}} analyzer to detect rule violations. Next, it employs predefined templates to transform the \emph{abstract syntax tree} (AST) of the program and synthesize patches that are well formatted, ready to review (can be merged into the code base without any change), and repair the violations. Thus, the burden on the developer is reduced from both interpreting and fixing rule violations, to inspecting and approving solutions for them. \revision{This can be of remarkable help to developers, as previous work shows the significant engineering effort required for keeping a project free of \sonar rule violations~\cite{saarimaki2019accuracy}.}

In order to make it even more convenient for developers to integrate \sorald into their development workflow, we also introduce \soraldbot. \soraldbot is a program repair bot \cite{van2019towards} that constantly monitors changes on \github repositories to find commits that introduce new violations. For such commits, \soraldbot utilizes \sorald to generate a patch that fixes the introduced violations and submits the patched version in the form of a pull request to the corresponding repository. The maintainers of the repository can then accept the pull request and merge it into their code base, or decline it. 

\revision{To study the applicability of \sorald, we run it on \topreposcnt distinct popular repositories on \github, which is a diverse datasets used for evaluating static violation repair tools.
In total, we detect 1,759 violations of the 10 \sonar rules that \sorald considers. \sorald targets a subset of 74\% (1,307/1,759) of these violations that meet repair preconditions. \sorald manages to fix 65\% (852/1,307) of the target violations, equal to 48\% (852/1,759) of all detected violations, which indicates the effectiveness of \sorald. We also let \soraldbot inspect the commit history of the \topreposcnt projects for 350 days. \soraldbot detects 46 commits in this period of time that introduce new violations of the considered \sonar rules in their changed files. \soraldbot generates 54 patches that fix the introduced violations. This shows the appropriate scale of contributions that \sorald can make, when integrated into popular development platforms like \github: it neither generates too many patches that would overwhelm projects maintainers, nor too few pull requests to make its contributions meaningless. Finally, we manually submit 29 pull requests based on patches generated by \sorald and analyze the reactions that we receive from the maintainers of the targeted projects.} Our analysis of these reactions provides us with insightful lessons that researchers can take into account to increase the acceptability of the contributions made by tools like \sorald. For example, we conclude that the contributions should be simple and short and focused on more severe violations.

Recently, a number of tools have been proposed to automatically fix issues detected by static analyzers~\cite{carvalho2020c,MarcilioFBP20,BaderSP019,phoenix}.
Compared to this recent related work, we focus on repairing violations labeled as \emph{bug} by the state-of-the-art static analyzer \sonarqube, which is one of the top static analyzers for Java used in industry \cite{vassallo2020developers}, with a bot that is integrated in the development workflow.

To summarise, we make the following contributions:
\begin{itemize}
    \item We introduce \sorald, a novel tool to automatically repair violations of static analysis rules, aka static analysis warnings, using generic transformations of the abstract syntax tree of the program under repair. 
    We implement our approach for Java for the popular static analyzer \sonar.
    
    \item \revision{We report original results of a large scale evaluation consisting of executing \sorald on \topreposcnt notable open source projects. The results show that \sorald fixes 65\% (852/1,307) of the violations which satisfy the repair preconditions.}
    
    \item \revision{We assess our proposed method in the context of a development workflow based on pull requests. By analyzing 350 days of the commit history of projects in our dataset, we show that \sorald is able to catch and repair 80 violations in those commits on the fly.}
    
    \item \revision{We perform a qualitative study based on developer feedback obtained after \sorald having generated 29 pull requests for 21 open source projects.} The accept/decline status of those pull requests, as well as the  written feedback from professional developers, indicate the appropriateness of \sorald contributions.
\end{itemize}

\textit{Structure of the Paper:} The rest of the paper is organized as follows: In \autoref{sec:sonar_background} we give a short background on \sonar. In \autoref{sec:approach}, we explain our proposed approach for automatically repairing static warnings. \autoref{sec:protocol} and \autoref{sec:results} present the protocol that we used for our experiments and their results. \autoref{sec:threats} discusses the threat to the validity of our results and also compares \sorald with the most similar tool, \spongebug. In \autoref{sec:related}, we review the related work. Finally, in \autoref{sec:conclusion} we conclude the paper.
\revision{This article is partially based on the master's theses of the 4th and 5th authors~\cite{adzemovic2020template,luong2021contributions}.}

\section{Analysis of \sonar}
\label{sec:sonar_background}

\sonar\footnote{\url{https://github.com/SonarSource/sonar-java}} is an open source static analyzer for Java code developed by \sonarsource, a company that specializes in code quality products\footnote{\url{https://www.sonarsource.com}}. These products are widely used by developers and have also garnered interest in the research community~\cite{LenarduzziS020,BaldassarreLRS20,MarcilioBMCL019,saarimaki2019accuracy}. In this section, we explain the \emph{rules} that \sonar bases its analysis on, and the \emph{rule violations} that arise when they are broken.

\subsection{Rules}
A rule in \sonar is a predicate that can be evaluated on a particular type of source code element, such as a method or a class. \revision{If the predicate is false, the rule is violated. Each rule is denoted with a unique identifier (SQID), a short one-sentence description of the rule, along with a more elaborate long description.}

As the purpose of \sonar is to alert developers about problems in the source code, most rules say what \emph{not} to do. For example, the short description of the rule with SQID \emph{S1217}\footnote{\url{https://rules.sonarsource.com/java/RSPEC-1217}} is ``\texttt{Thread.run()} should not be called directly'', and the long description explains that this causes the code to be executed in the calling thread instead of a new thread, as is most likely the intention. This rule can be considered as a predicate over method calls that evaluates to true if the receiver is of type \texttt{Thread}, and the called method's signature is \texttt{run()}.

In many cases, \sonar also suggests how a developer can address the problem. In the case of rule \emph{S1217}, the clear-cut solution suggested in the documentation is to call \texttt{Thread.start()} instead, which has the desired effect of running the code in a separately created thread. It is however up to the developer to actually implement the fix.

Some rules deviate from the usual pattern and tell the developer what to do, rather than what not to do. For example, rule \emph{S2095}\footnote{\url{https://rules.sonarsource.com/java/RSPEC-2095}} has the short description ``Resources should be closed'', and is raised on resources such as file streams that are opened but potentially not closed. In this case the rule is broken if the predicate is false, and the high-level solution to the problem is directly embedded in the rule's phrasing.

\subsection{Violations}
\label{sec:sonar_violations}

\begin{lstlisting}[float=t, style=diff, caption={\revision{Example of a fix for a violation of \textit{S1217}. \sorald is able to generate this fix.}}, captionpos=b, label=lst:transform_ex]
  Runnable runnable = () -> System.out.println("Hello world!");
  Thread myThread = new Thread(runnable);
%\RHilight%- myThread.run();
%\GHilight+ myThread.start();
\end{lstlisting}

\revision{In this paper, we define violations of \sonar rules and their fixes as follows.}

\revision{\textbf{Definition} A \emph{rule violation} is a code fragment that violates a predicate dictated by a rule and implemented in \sonar.}

\revision{When there is a rule violation, \sonar reports it by stating the ID of the violated rule and the position of the violating code fragment. We use the term rule violation both to refer to the problematic code fragment and to the warning reported by \sonar, the distinction between these two is not significant in the context of this paper.}

\revision{\textbf{Definition} A \emph{fix} (or \textit{repair}) for a rule violation $v$ is a patch that edits the source code such that the patched code satisfies two conditions. First, \sonar does not report $v$ when it scans the patched source code. Second, the patched code still implements the behavior expected by the developer.}

\revision{Consider an example of a rule violation and its fix in \autoref{lst:transform_ex}. The original code in this listing (lines 1, 2, and 3) shows an example code snippet where rule \emph{S1217} is violated. \sonar reports a rule violation stating that \texttt{Thread.run()} should not be called on line 3.}

\revision{In isolation, fixing this single violation is not a problem for any developer, as it is a matter of replacing \texttt{run} with \texttt{start}, given that this modification produces the expected behavior. However, this is
\begin{enumerate*}
    \item just a single violation of this rule, and there could be many more in the project, and
    \item just a single rule out of \sonar's 631 rules\footnote{\url{https://rules.sonarsource.com/java}, visited on Nov 21, 2021.}
\end{enumerate*}.
Consequently the value of automatically fixing those violations does not necessarily come from relieving the developer of a difficult task, but rather of helping to cope with a multitude of chores.
The engineering effort of keeping a project free of \sonar rule violations, or even just preventing new ones from being introduced, is significant \cite{saarimaki2019accuracy}.}

\revision{Importantly, an expected behavior that is already implemented should not be broken after a violation fix, i.e., a fix should not introduce any regression.
However, after fixing a violation, the program behavior may slightly change, but only from a non-functional perspective or for corner-case inputs, which is the aim of \sonar rules.
For example, \autoref{lst:transform_ex} shows a fix for a violation of rule \textit{S1217}. This piece of code is expected to execute the \texttt{runnable}, defined at line 1, in a separate thread. Line 3 of the original code is an incorrect implementation of this expected behavior, as it executes the \texttt{runnable} in the current thread. This violates rule \textit{S1217}, which is designed to detect this type of incorrect implementation. The fix for this violation modifies the invocation of \texttt{myThread.run()} method at line 3 to the invocation of \texttt{myThread.start()} at line 4 to execute the \texttt{runnable} in a separate thread. After applying this fix, \sonar no longer detects a violation of \textit{S1217} and the expected behavior is correctly implemented. To summarise, the parts of the expected behavior that are implemented correctly in the original code, such as defining the \texttt{runnable} at line 1, are left unchanged, no regression is introduced and the rule violation has disappeared.}

\section{The \sorald Approach}
\label{sec:approach}

\begin{figure*}
\begin{center}
\includegraphics[width=\textwidth]{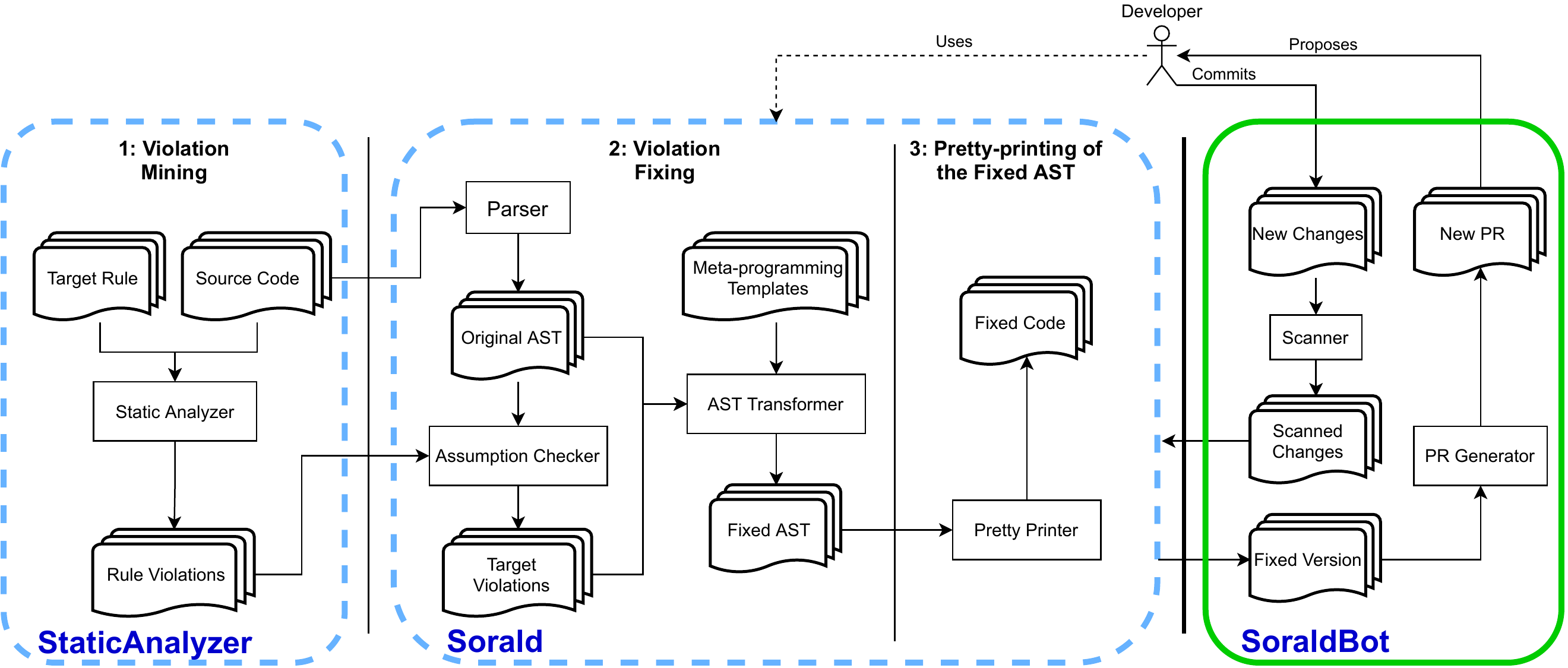}
\caption{\revision{Overview of the \sorald approach for repairing \sonar static analysis violations.}}
\label{fig:sorald}
\end{center}
\end{figure*}

\sorald is a novel approach to detect and fix violations of static analysis rules. The violations are repaired at the level of the \emph{abstract syntax tree} (AST) of the program using metaprogramming.

\subsection{Overview}
\autoref{fig:sorald} shows the workflow of our approach, which consists of three steps, the two latter ones being \sorald itself.
First, \sonar performs \emph{violation mining}, to detect rule violations. Second, \sorald performs \emph{violation fixing}, where it repairs each violation by transforming the AST of the program using metaprogramming templates. Third, and finally, \sorald performs \emph{pretty-printing} to turn the AST back into source code.

As shown in \autoref{fig:sorald}, the second and third steps comprise the core of \sorald, while the first step of the repair process is mostly conducted using \sonar.
We describe these steps in more detail in the following sections.

\subsection{Violation Mining}
\revision{\sorald takes the source code of a program and a target rule that the user wishes to fix violations of as input. \sorald then employs a static analyzer to statically analyze the code and find all rule violations and their exact positions in the source code. Currently, \sorald uses \sonar as the static analyzer that identifies rule violations. Nonetheless, \sonar can be replaced by any other tool that can detect the exact location of violations. These violations are then passed to the violation-fixing step, which is described next.}

\subsection{Violation Fixing}
\label{sec:violation_fixing}

In the second step, \sorald fixes rule violations found during mining. As mentioned in \autoref{sec:approach}, \sorald performs repairs on the AST of the program. Therefore, it first parses the source code to obtain the original AST. The AST and rule violations are passed to an \emph{Assumption Checker}, which matches the violations to AST elements based on source code positions and verifies that \sorald can repair them, discarding any violations that it cannot handle. This check is necessary as some \sonar rules are broadly defined, and need to be further qualified. Consequently, \sorald only targets a subset of violations of these rules, which are refined with assumptions. 

For example, rule \textit{S2225} dictates that methods \texttt{toString()} and \texttt{clone()} should not return \texttt{null}, but these two different cases require radically different repair approaches. Fixing \texttt{null}-returning instances of \texttt{toString()} is an appropriate task for template-based repair, as it is a matter of returning the empty string instead of \texttt{null}. However, \texttt{clone()} returns instances of the declaring class, which may not be possible to synthesize, and the general case is therefore not amenable to template-based repair. Therefore, \sorald only targets \texttt{null}-returning \texttt{toString()} methods for this rule, and violations on \texttt{clone()} are discarded by the assumption checker.
As of now, we refer to violations that pass the assumption checker as \emph{\target violations}.

After finding \target violations, they are passed along with the original AST to the \emph{AST Transformer}. The transformer selects the appropriate templates to apply, and applies them to the corresponding AST elements. For each violation, it applies one metaprogramming template to the corresponding AST element to fix it. 

\autoref{lst:transform_ex} shows the textual \emph{diff} of one such template fixing the violation of rule \emph{S1217} discussed in \autoref{sec:sonar_violations}. Lines starting with \texttt{+} are added with the fix, and lines starting with \texttt{-} are removed. Recall that \emph{S1217} dictates that \texttt{Thread.run()} should not be called directly, and that \texttt{Thread.start()} should be used instead. This is precisely the transformation applied in \autoref{lst:transform_ex}.

\autoref{lst:processor_ex} shows how the corresponding metaprogramming template is implemented. The input is an AST element of type \texttt{CtInvocation} (a method call). This element has already been identified to violate rule \textit{S1217} by calling the \texttt{run()} method of a \texttt{Thread} object. The template fixes the violation by creating an invocation of the \texttt{start()} method on the same object and replacing the input element with the newly created invocation element.

Note that the template shown in \autoref{lst:processor_ex} is among the most simple ones, with a single code path for fixing the problem. This does not mean that one template is equivalent to one fix; templates can adapt to the context of the violating element and apply different fixes for different violations of the same rule. For example, rule \emph{S4973}\footnote{\url{https://rules.sonarsource.com/java/RSPEC-4973}} states that ``Strings and Boxed types should be compared using \texttt{equals()}'', as opposed to \texttt{==} or \texttt{!=}. As this template handles expressions both for equality and inequality, it must replace usages of the mentioned operators with calls to \texttt{equals()}, but also take care to prepend a \texttt{!} to the resulting expression if \texttt{!=} is used in the original comparison. This is a simple example of a context-sensitive template.

When all \target violations have been processed by templates, the fixed AST is passed on to the pretty-printing step.

\begin{lstlisting}[float=tb, style=diff, caption={Excerpt of the code that transforms the AST of a program to fix violations of \sonar's \textit{S1217} rule.}, captionpos=b, label=lst:processor_ex]
void transform(CtInvocation<?> element) {
  Factory factory = element.getFactory();
  CtClass<?> threadClass = factory.Class().get(Thread.class);
  CtMethod<?> method = threadClass.getMethodsByName("start").get(0);
  CtInvocation<?> threadStartInvocation =
  factory.createInvocation(element.getTarget(), method.getReference());
  element.replace(threadStartInvocation);
}
\end{lstlisting}

\subsection{High-Fidelity Pretty-Printing}

In the final step, \sorald converts the fixed ASTs into Java source code by means of pretty-printing. 
To be useful, the user of \sorald needs to spend less time on reviewing and merging patches than it would take for a developer to manually fix the violations. To achieve this goal, \sorald needs to produce patches that preserve the \textit{documentary structure}, that is the formatting and comments of the original sources\cite{DeVanter2001}, as much as possible. Failing to do so results in patches that are both (i) less readable, as formatting is crucial for source code readability~\cite{buse2008metric}, and (ii) are more time-consuming to review, as they would cause larger diffs.

Tools for AST-based transformation typically discard documentary structure during parsing, as it does not contain semantic information. For many applications, static analyzers included, this does not pose a problem as there is no need to convert the AST back to source code. When there is a need to go back to source code, finding a way to accurately preserve documentary structure after transforming an AST is challenging, and is known as the problem of high-fidelity pretty-printing \cite{kort2003}.
To address this issue, most tools represent the documentary structure as nodes or annotations attached to previous or leading AST nodes \cite{Kitlei2009,kort2003} but as De Jonge et al.\cite{deJonge2012} explain it, this approach has several shortcomings since it is not necessarily possible to unambiguously decide which concrete AST node should retain which parts of the documentary structure. Indeed, comments may relate to several nodes, and the same whitespace can trail one node, or lead another, which becomes problematic at the edges of node lists.

To meet the requirements of high-fidelity pretty-printing, we implement the technique of De Jonge et al.\cite{deJonge2012} into our AST transformation library.
In \sorald, with the underlying \spoon library \cite{spoon}, we build a Source Fragment Tree that is separated from the AST. When pretty-printing the fixed ASTs, we match unchanged AST nodes with the corresponding source fragment from the original sources, and print it. We only rely on a standard pretty-printer when we fail to match a node to a source fragment or when the node has been changed by a transformation.
By using this approach, we ensure that \sorald minimizes changes to documentary structure such as whitespace, comment placement and formatting, as well as parenthesization. In doing so, \sorald maximizes the clarity of patches suggested to developers.

\subsection{Considered Rules}
\label{sec:considred_rules}

\revision{Here we explain how we determine which \sonar rules are considered in \sorald.}

\subsubsection{\revision{Fixability Analysis}}
\label{sec:rule_taxonomy}

\begin{table}[t]
\centering
\small
\caption{\revision{Comprehensive Manual Analysis of the 631 \sonar Rules.}}
\label{tab:sonar_rule_types}
\begin{tabular}{ l r r r}
\toprule
     & & \multicolumn{2}{c}{\footnotesize Prevalent in real-world?} \\
	\textbf{Rule Type}  & \textbf{\#Rules} & Yes & No \\
	\midrule
	Bug rules & 153 (25\%) & -- & -- \\
    \quad Fully fixable rules & 77 (13\%) & 27 (5\%) & 50 (8\%) \\
    \quad Partially fixable rules  & 20 (3\%) & 9 (1\%) & 11 (2\%) \\
    \quad Unfixable rules & 56 (9\%) & 17 (3\%) & 39 (6\%) \\
    \bottomrule
\end{tabular}
\end{table}

\revision{To decide what rules should be considered in \sorald, we conduct a deep analysis of all 631 rules in \sonar to detect the ones whose violations can be fixed with our approach. In this manual analysis, we extract two features for each rule.
\begin{enumerate*}
    \item Rule type: is it a ``bug'' rule? 
    \item Fixability: is the rule fixable with our template-based AST transformations?
\end{enumerate*}
}

\revision{The \emph{rule type} filter yields 153 bug rules. To find out the fixability of those 153 rules, we divide them into three groups as follows.}

\revision{\textbf{1- Fixable:} We label a rule $r$ as \emph{fixable} if it is possible to design a metaprogramming template that meets the following conditions. First, after the template is applied on the program, \sonar no longer reports $r$. Second, after the template is applied on the source code, a single patch is generated. Third, the patch implements the expected behavior under the most common understanding of the code under consideration.}

\revision{\emph{Single patch for a given violation.} Regarding the second condition, metaprogramming templates in \sorald are fully deterministic. This means a \sorald template is meant to generate one and only one patch for a given rule violation, and that patch is always the same. For some \sonar rules, there are multiple ways of violating the rule. Each case requires a specific type of change on the code to be repaired. When static analysis cannot determine what type of change is required for every given violation of a rule, it is not possible to design a template to fix all violations of that rule. For example, \textit{S1221} is a rule that dictates that methods should not be named ``\texttt{tostring()}'' (with lowercase 's')~\footnote{\revision{\url{https://rules.sonarsource.com/java/RSPEC-1221}}}. Two types of errors can lead to violating this rule: misspelling capital 'S', when the intention is to override the \texttt{toString()} method, or purposefully using lowercase 's', when the intention is to not override \texttt{toString()}. Both types of mistakes are possible but it cannot be statically determined which scenario a given violation belongs to. Therefore, we cannot design a template that automatically fixes all violations of \textit{S1221} with a single patch.}

\revision{\emph{Most common understanding.} With respect to the third condition, we consider that \sonar documents explain the common understanding of a rule violating program. When these reference documents are not clear wrt the likely intention of the program, it is not possible to design a template in \sorald. For example, the documentation for rule \textit{S2583}~\footnote{\revision{\url{https://rules.sonarsource.com/java/RSPEC-S2583}}} notes that every branch in a program should be reachable. However, the document does not state what should be done with unreachable code (reach it or remove it?). Therefore, the expected behavior of a program violating this rule cannot be determined under a common understanding and it makes it impossible to design a repair template. Consequently, rule \textit{S2583} cannot be considered to be fixable.}

\revision{\textbf{2- Partially fixable:} A rule is \emph{partially fixable} if and only if 1) a strict subset of its violations are fixable and the remaining are not fixable and 2) it is possible to statically determine whether or not a violation lies in the fixable subset. For example, as discussed in \autoref{sec:violation_fixing}, a subset of \textit{S2225} violations that occur in the \texttt{toString()} method are fixable, as they can be fixed by replacing \texttt{return null} with \texttt{return ""}. On the other hand, the violations of \textit{S2225} that occur in the \texttt{clone()} method are not fixable, as we cannot guess what a \texttt{clone()} function should return instead of \texttt{null} to implement the expected behavior. For this reason, we label rule \textit{S2225} as partially fixable.}

\revision{\textbf{3- Unfixable:} The rules that are neither fixable nor partially fixable are labeled as \emph{unfixable}.}

\revision{To ensure accurate labeling for rules as fixable/partially fixable/unfixable, two authors (a PhD student and a Postdoc researcher) conduct this analysis in parallel and resolve conflicting labels with thorough discussions.}

\revision{\autoref{tab:sonar_rule_types} presents the results of our analysis on \sonar rules. Out of 631 rules, 478 (75\%) are non-bug rules, which means they are not in the scope of \sorald repairs. Among the remaining 25\% (153 out of 631) bug rules, we label 56 of them as unfixable. 20 (3\% of 631) and 77 (13\% of 642) are partially fixable and fixable, respectively. This shows that metaprogramming with templates is appropriate to fix violations of \sonar bug rules.}

\subsubsection{Selecting the Rules to Implement}
\label{sec:rule_selection}

\begin{table*}[t]
\centering
\footnotesize
\begin{threeparttable}
\caption{\revision{The 10 \sonar static analysis rules considered in this study, all labeled with an importance level of \emph{BUG} and implemented in \sorald.}}
\label{tab:rules}
\begin{tabular}{| m{0.7cm} | m{5.7cm} | m{8.6cm} | m{1.5cm} |}
\hline
	\textbf{SQID} & \textbf{\sonar Description} & \textbf{Fix Template} & \textbf{Human remediation time\tnote{1}} 
	\\ \hline \hline
	
	S1217 & \texttt{Thread.run()} should not be called directly.                        & 
	\revision{\sorald fixes the violations of this rule by replacing each invocation of \texttt{Thread.run()} with an invocation of \texttt{Thread.start()}. Implementation: \href{https://bit.ly/3pLhYZb}{bit.ly/3pLhYZb}}
	& 20 min 
	\\ \hline

	S1860 & Synchronization should not be based on Strings or boxed primitives.    & 
	\revision{If the lock is a field of the current class where the synchronization block is in, then \sorald adds a new field as an \texttt{Object} lock. If the lock is obtained from another object through the \texttt{get} method, \sorald will add a new field for the new \texttt{Object} lock and a new method to get the object. Implementation: \href{https://bit.ly/3IyINs2}{bit.ly/3IyINs2}}
	& 15 min
	\\ \hline 

	S2095 & Resources should be closed.                                            & 
	\revision{The repair encloses the parent block of resource initialization in a try-with-resources. If it was already in a try block, \sorald replaces the try with try-with-resources instead of creating a new one, so that useless nested try blocks are not created. Implementation: \href{https://bit.ly/3lPe8wP}{bit.ly/3lPe8wP}}
	& 5 min 
	\\ \hline 

	S2111 & \texttt{BigDecimal(double)} should not be used.                             & 
	\revision{The constructor of \texttt{BigDecimal} is replaced with \texttt{BigDecimal.valueOf(parameter)} if the constructor has one argument, otherwise, the first parameter is enclosed in a \texttt{String}. Implementation: \href{https://bit.ly/3dyoa0X}{bit.ly/3dyoa0X}}

	& 5 min
	\\ \hline

	S2116 & \texttt{hashCode} and \texttt{toString} should not be called on array instances. & 
	\revision{Any invocation of \texttt{toString()} or \texttt{hashCode()} on an array is replaced with \texttt{Arrays.toString(parameter)} or \texttt{Arrays.hashCode(parameter)}. Implementation: \href{https://bit.ly/3oCmSrN}{bit.ly/3oCmSrN}}
	& 5 min
	\\ \hline 

	S2142 & \texttt{InterruptedException} should not be ignored.                        & 
	\revision{A catch block that catches an \texttt{InterruptedException}, but neither re-interrupts the method nor rethrows the \texttt{InterruptedException}, i.e., ignores the \texttt{InterruptedException}, is augmented with \texttt{Thread.currentThread().interrupt();}. Implementation: \href{https://bit.ly/3IyJq4S}{bit.ly/3IyJq4S}}
	& 15 min 
	\\ \hline 

	S2184 & Math operands should be cast before assignment.                        & 
	\revision{In arithmetic expressions, when the operands are \texttt{int} and/or \texttt{long}, but the result of the expression is assigned to a \texttt{long}, \texttt{double}, or \texttt{float}, the first left-hand operand is cast to the final type before the operation takes place. To the extent possible, literal suffixes (such as \texttt{f} for \texttt{float}) are used instead of casting literals. Implementation: \href{https://bit.ly/31zTbPH}{bit.ly/31zTbPH}}
	& 5 min
	\\ \hline 

	S2225 & \texttt{toString()} and \texttt{clone()} methods should not return null.         & 
	\revision{For the return statements inside \texttt{toString()}, this processor replaces the return expression with an empty string. 
    Note that this processor is partial (partially fixable rule) and does not fix null-returning \texttt{clone()} methods. Implementation: \href{https://bit.ly/31zTkCJ}{bit.ly/31zTkCJ}}
	& 5 min
	\\ \hline 

	S2272 & \texttt{Iterator.next()} methods should throw \texttt{NoSuchElementException}.   & 
	\revision{Any implementation of the \texttt{Iterator.next()} method that does not throw \texttt{NoSuchElementException} has a code snippet added to its start. The code snippet consists of a call to \texttt{hasNext()} and an appropriate throw statement. Implementation: \href{https://bit.ly/3EMCn6s}{bit.ly/3EMCn6s}}
	& 5 min
	\\ \hline 

	S4973 & Strings and Boxed types should be compared using \texttt{equals()}.         & 
	\revision{Any comparison of strings or boxed types using \texttt{==} or \texttt{!=} is replaced with \texttt{equals()}. Implementation: \href{https://bit.ly/3DG77Vh}{bit.ly/3DG77Vh}}
	& 5 min
	\\ \hline 
\end{tabular}
\begin{tablenotes}
    \item[1] Per \sonar itself
\end{tablenotes}
\end{threeparttable}
\end{table*}

We design \sorald to detect and fix violations for the most notable \sonar rules. We select the rules to be implemented according to the following criteria:
\begin{enumerate*}
    \item \revision{The type of the rule should be \emph{bug}.}
    \item \revision{The rule should be fixable or partially fixable as per the analysis in \autoref{sec:rule_taxonomy}. This criterion ensures that we can come up with a proper template.}
    \item The rule should have at least one violation reported in Marcilio et al.'s study \cite{MarcilioBMCL019}. In this paper, Marcilio et al. provide a dataset of 421,976 \sonar issues from 246 Java projects. This criterion guarantees that fixing violations of the selected rules is useful for the practitioner community.
\end{enumerate*}

\revision{All the information required for selecting the rules is shown in \autoref{tab:sonar_rule_types}. The last two columns, ``Yes'' and ``No'' show the number of rules with and without an instance in Marcilio et al.'s dataset~\cite{MarcilioBMCL019}, respectively. There are 36 (9 partially fixable and 27 fixable) rules that meet all the criteria mentioned above. To implement a first version of our proposed repair technique in \sorald, we select the 10 rules that: 1) are among the most frequent, and 2) their automatic fixing templates can be implemented in a reasonable amount of engineering effort and time\footnote{The design and implementation of \sorald already represents 2+ years of full time work. Note that commercial tools, such as \sonar, have large teams of tens of developers to design and implement rule based checks to detect potential bugs in programs. Fixing violations of all of those rules is out of reach for a comparatively small academic research group.}}

\revision{\autoref{tab:rules} presents the 10 selected rules. The first two columns present the unique identifier and the short description of the rules, as described in \autoref{sec:sonar_background}. The third column shows the template that \sorald uses to fix violations of the corresponding rule. This column also proves a link to \sorald's implementation of these templates in the Java language. Finally, ``RT'' is the \emph{remediation time} in minutes, which is an estimate of the amount of time required to manually fix the violation. The data related to the remediation time is taken directly from the \sonar repository\footnote{\url{https://github.com/SonarSource/sonar-java/tree/970d69bbeb741abed1ef65a1e641a847af9bf0e3/java-checks/src/main/resources/org/sonar/l10n/java/rules/java}}.}

\revision{For example, rule \textit{S4973} indicates that strings and boxed types should be compared with \texttt{equals()}. To fix violations of this rule, when two strings or boxed type objects are compared with \texttt{==} or \texttt{!=}, \sorald replaces the operator with \texttt{equals()}. This is a direct fix for violations of \textit{S4973} that always removes the violation when applied.}

\subsection{Development Workflow Integration}
\label{sec:integration-workflow}

In its simplest form, \sorald can be used as a stand-alone tool that is executed from the command-line to repair violations in a target program. On the other hand, it is also designed to be integrated into popular development workflows in order to fully automate \sorald's patch suggestions. \sorald offers the latter option with an integration service called \emph{\soraldbot}. \soraldbot is a program repair bot \cite{van2019towards} that automatically \emph{scans} changes in a set of target repositories, \emph{identifies} changes that introduce new violations, \emph{fixes} the new violations, and \emph{proposes} the fixed version to the developers as a patch. As \soraldbot is designed to operate on \github, it proposes the patch in the form of a \emph{pull request} (PR)\footnote{\url{https://docs.github.com/en/github/collaborating-with-issues-and-pull-requests/about-pull-requests}}, which represents a request by a contributor to merge a patch into a repository on the \github platform. A maintainer of a repository can easily review the patch right on the \github platform, potentially request further changes, and then merge the patch into the repository with the click of a button. Crucially, this workflow allows a developer (or bot) that does not have write access to the targeted repository to easily propose patches.

\autoref{fig:sorald} shows an overview of how \soraldbot works (the box with solid borderlines). \soraldbot's scanner constantly monitors target repositories. Once a developer commits new changes, \soraldbot analyzes those changes. Next, \soraldbot takes advantage of the same part of \sorald that is executed from command-line to mine the changed files. If any new violations that \sorald is equipped to handle are introduced, \soraldbot uses \sorald to repair the files and obtain a \emph{fixed version}. Finally, a pull request generator creates a \emph{new PR} based on the fixed version and proposes it to the developer.

At this point, a maintainer of the targeted repository decides whether to merge the PR. We refer to a merged PR as \emph{accepted}, and a PR that is closed without merge as \emph{declined}.

\revision{The originality of \soraldbot is that it fixes the violations in an integrated manner: it monitors open source projects, detects static rule violations, produces fixes, and suggests them to the developers. Considering that many static warnings are ignored by developers due to the hardship of fixing them \cite{MarcilioBMCL019}, our software bot can help developers to take better advantage of static analysis tools. It is a pragmatic approach to static analysis; developers are only alerted of a problem if it can be automatically fixed.}

\subsection{Implementation}
\sorald is implemented in the Java programming language.
It uses the \sonar static analyzer to find rule violations, and leverages the \spoon\cite{spoon} metaprogramming library to parse source code, perform AST transformations, and convert the fixed AST back to Java source code with high-fidelity pretty-printing. 

\revision{\sorald is extensible and is not bound to only use \sonar. The only requirement from \sorald is that the static analyzer in question provides an accurate source code position for the violation. Hence, rule based static analyzers like SpotBugs \cite{lavazza2020empirical} and PMD \cite{trautsch2020longitudinal} can substitute \sonar. Additionally, even if no such static analyzer is provided, \spoon has all the facilities necessary for performing static analysis, meaning that rule violation detection can be implemented directly in \sorald.}

\sorald's source code and all relevant experimental data is made publicly available for future research on this topic. The repository can be accessed at: 
\begin{center}
    \ULurl{https://github.com/SpoonLabs/sorald}
\end{center}


\section{Experimental Protocol}
\label{sec:protocol}
We now present our systematic experimental protocol to evaluate \sorald.

\subsection{Research Questions}

\newcommand\rqone{How applicable is \sorald on a diverse set of real-world projects?}

\newcommand\rqtwo{What is the intensity of \sorald's activity if integrated into a development workflow?}

\newcommand\rqthree{What is the opinion of developers about \sorald fix suggestions?}

In this paper, we study the following research questions.

\begin{itemize}
    \item RQ1: \rqone \ To assess the applicability of \sorald, we measure its accuracy in terms of detecting and fixing violations of \sonar rules on a large and diverse set of open source projects.
    
    \item RQ2: \rqtwo \ To evaluate the potential effect of integrating \sorald into continuous integration workflows, we measure the number of violation-fixing patches that \soraldbot creates in a certain period of time.
    
    \item RQ3: \rqthree \ To investigate developers' reactions to \sorald fix suggestions, we create real pull requests on active open source projects. We target a curated set of projects and study the reactions of their maintainers to the created pull requests.
\end{itemize}

\subsection{Study Subjects}
\label{sec:dataset}
We conduct our experiments on a curated collection of \github projects. We call this curated list \toprepos.
\toprepos includes all Java repositories on \github that are:
\begin{enumerate*}
    \item Popular project: have at least 50 stars on \github.
    \item Maven project: have a ``pom.xml'' file in the root directory.
    \item Active project: have at least one commit in the past three months as of Nov 2020.
    \item Pull request friendly project: have a pull request accepted in the last three months as of Nov 2020.
    \item Healthy project: pass ``mvn compile'' and ``mvn test'' commands successfully with OpenJDK Java 11 on the last commit in Nov 2020. 
    \item Continuous integration friendly project: the project uses CI, which is checked as having a ``.travis.yml'' in the root directory.
\end{enumerate*}
This dataset includes \topreposcnt repositories in total and it is publicly available in our open science repository~\footnote{\url{https://github.com/SpoonLabs/sorald}}.

\begin{table}[t]
\centering
\caption{Characteristics of the \topreposcnt repositories considered as study subjects.}
\label{tab:dataset_stats}
\begin{tabular}{@{}l r r r r r r@{}}
\toprule
	{} & Min & Q1 & Median & Q3 & Max & Sum\\
	\midrule
	\#Violations & 0 & 7 & 21 & 70 & 1,652 & 14,842 \\
	\#KLOC & 0.04 & 3.26 & 9.73 & 23.02 & 502.44 & 4,110.33 \\
	\#Contributors & 2 & 6 & 11 & 23 & 150 & 3,269 \\
	Age (months) & 9 & 50 & 77 & 95 & 147 & 12,015 \\
	\#Commits & 16 & 125 & 254 & 623 & 11,347 & 119,127 \\ \hline
	\bottomrule
\end{tabular}
\end{table}

\autoref{tab:dataset_stats} describes this dataset of \topreposcnt repositories. The table shows the minimum, median, and maximum number of violations of importance `bug' in the selected repositories. ``Q1'' and ``Q3'' stand for the first and third quartiles, respectively. The last column (``Sum'') shows the total number of violations in the dataset. To give a sense of repositories size, it also presents statistics on their number of lines of code, contributors, commits, and the age.
\revision{As shown in the table, the median number of lines is 9.73 thousand and the median number of commits is 254. This shows that most of the included projects are of notable size.}
The median number of violations is 21 which indicate that \sorald is effective at identifying problems in those projects.

\subsection{Methods}
\subsubsection{RQ1}
\label{sec:protocol1}

To answer \textbf{RQ1}, we run \sorald on the latest version of projects included in \toprepos, and for each of them we use \sorald to repair the \sonar violations. We count the number of \emph{}{``\target''} violations and all detected violations for each considered rule before and after running \sorald. ``TDR'', as defined in \eqref{eq:target_to_detected}, shows the ratio of \target violations to all detected violations (see \autoref{sec:violation_fixing} for description of \target violations). If \textit{TDR} is close to $1.0$ for a rule, it indicates that most violations of that rule are of the type that \sorald attempts to repair.

After computing the raw numbers, we compute two ratios for each considered rule to assess the applicability: the ratio of \target violations that are fixed and the ratio of all detected violations that are fixed. These ratios are calculated according to \eqref{eq:starget_fixed} and \eqref{eq:detected_fixed}, respectively. A high ``FTR'' means \sorald is able to fix violations of its target type, while ``FDR'' shows the effectiveness of \sorald on violations of considered rules regardless of whether they are of the target type or not.

\begin{equation}
\label{eq:target_to_detected}
\resizebox{0.5\hsize}{!}{$\mathit{TDR}=\frac{|Target\ Before\ Repair|}{|All\_Detected\ Before\ Repair|}$}
\end{equation}

\begin{equation}
\footnotesize
\label{eq:starget_fixed}
\resizebox{1\hsize}{!}{$\mathit{FTR}=\frac{|All\_Detected\ Before\ Repair|-|All\_Detected\ After\ Repair|}{|Target\ Before\ Repair|}$}
\end{equation}

\begin{equation}
\footnotesize
\label{eq:detected_fixed}
\resizebox{1\hsize}{!}{$\mathit{FDR}=\frac{|All\_Detected\ Before\ Repair|-|All\_Detected\ After\ Repair|}{|All\_Detected\ Before\ Repair|}$}
\end{equation}

In \eqref{eq:detected_fixed} and \eqref{eq:starget_fixed}, \emph{FDR} and \emph{FTR} represent the ratio of fixed all detected violations and fixed \target violations, respectively. Also, note that the numerator in these two equations (\resizebox{1.0\hsize}{!}{$|All\_Detected\ Before\ Repair|-|All\_Detected\ After\ Repair|$}) is equal to the number of \emph{fixed violations}.

After computing the number of fixed violations, we also calculate the estimated remediation time for fixed violations. We simply obtain this value by multiplying the number of fixed violations by the estimated remediation time for one violation of the corresponding rule (according to ``RT'' in \autoref{tab:rules}). This metric gives us a rough sense of the value of \sorald repairs.

Moreover, we compile the projects and execute their tests after they are patched by \sorald. This tells us if \sorald changes the tested behavior, and consequently whether it can be trusted to not introduce regressions after applying the fixes. We calculate the ratio of projects with failing tests after repair to evaluate to what extent \sorald can be trusted. \revision{Since it is possible to have flaky tests that fail for reasons other than \sorald patches, we manually analyze the results to determine whether the \sorald patches are the cause of the failure.}

\revision{In addition to this per rule analysis, we also compute the total number of violations of all fixable and partially fixable bug rules of \sonar in our dataset: there are 77 fixable and 20 partially fixable bug rules in \sonar, which makes the total number of such rules 97. Based on the total number of violations of these rules, we also compute the ratio of all bug related violations that are covered by our considered rules. We calculate this ratio by dividing the number of all detected violations of our 10 considered rules by the total number of all violations of 97 fixable and partially fixable rules. This ratio suggests how significant and representative our considered rules are.}

Finally, to appraise \sorald's performance, we measure the time it spends to fix violations of a rule in a specific project. We perform this assessment while running our experiment on a machine with \revision{sixteen Intel(R) Core(TM) i9-10980XE processors, each running at 3.00GHz and having 18 cores, and eight 16GB RAMs each of type DDR4 with 3600 MT/s clock speed}. Then, for each rule, we compute the median time spent to fix its violations on a single project.

\subsubsection{RQ2}
\label{sec:protocol2}
\revision{In this experiment, we aim to know the potential scale of contributions that \soraldbot can make. To answer \textbf{RQ2}, we count the number of patches that \soraldbot would create on the projects of \toprepos over a period of 350 days. We set this time period such that we can conduct the whole experiment in five days. We inspect the commits in retrospect. For this purpose, on Nov 27, 2021, we run \soraldbot to analyze all commits in the \toprepos projects from Dec 12, 2020 to Nov 27, 2021.} We set up \soraldbot to generate a separate patch for each rule \textit{r} that is violated in the changed files of a commit \textit{c}. This patch contains fixes for all violations of \textit{r} in \textit{c}. We do not submit the generated patches as pull requests to the corresponding repositories because the commits are too old to be considered as active work items.

\revision{The key metrics that we measure in this experiment are the number  violation introducing commits, the number of projects including these commits, the number of fixed violations in the violation introducing commits, and the number of generated patches.} We break the metrics per \sonar rules. These metrics indicate the scale of the code contributions that \sorald can potentially make to a repository if integrated as an automated software bot.


\subsubsection{RQ3}
\label{sec:protocol3}


\revision{To answer \textbf{RQ3}, based on the information gathered in the RQ2 experiment, we submit a number of pull requests to fix  violations in \toprepos projects. Suppose \textit{fs} is the set of files of a project \textit{p} that are changed between Dec 12, 2020 and Nov 27, 2021 and also introduce new violations of a considered rule \textit{r}. As described in \autoref{sec:protocol2}, in the RQ2 experiment, we use \soraldbot to detect such changes and produce patches that fix the introduced violations. Each of these patches correspond to a commit that is introducing violations of \textit{r}. In our experiment to answer RQ3, we aggregate all the patches from the RQ2 experiment that fix violations of \textit{r} in commits of \textit{p} in the studied period. Next, we manually submit a pull request to \textit{p} that applies all the aggregated patches on the latest version of \textit{fs}. This means each pull request fixes violations of one single rule. Some patches correspond to old commits and the patched code does not exist anymore. We ignore those patches, as they cannot be applied on the latest version of \textit{fs}. We focus on recently changed filed from the RQ2 experiment. This makes it more likely that the maintainers of \textit{p} have the context of violations in mind. After submitting the pull requests, we follow up with the maintainers if they ask questions or leave comments.}


In this experiment, we assess the acceptability of \sorald contributions and potential paths to improve it. For this purpose, we perform a qualitative analysis of the reactions received from developers. Besides looking at their decisions (accept/decline), we engage in a discussion with project maintainers to see what they think about the idea of a software bot aimed at fixing rule violations and what they see as possible improvements in \sorald. 


\section{Experimental Results}
\label{sec:results}
We now present our experimental results in answer to our research questions.

\subsection{Applicability of \sorald (RQ1)}
\label{sec:rq1_results}

\begin{table*}[t]
\centering
\setlength{\tabcolsep}{1em}
\caption{\revision{The applicability of \sorald on 161 open source projects. \sorald successfully repairs a majority of the considered violations (65\%) in the order of magnitude of seconds per violation, which saves valuable time of human developers.}}
\label{tab:ex1_res}
\begin{threeparttable}[b]
\begin{tabular}{l | r r r r r r}
\toprule
	\textbf{SQID} & \textbf{TDR (TV/DV)} & \textbf{FTR (FV/TV)} & \textbf{FDR (FV/DV)} & \textbf{TRT (min)} & \textbf{Failing\_Repos} & \textbf{MT (sec)} \\
	\midrule
    S1217 & 100\% \hfill (2/2) & 100\% \hfill (2/2) & 100\% \hfill (2/2) & 40 & 0.0\% (0/161) & 4.5 \\
    S1860 & 100\% \hfill (5/5) & 100\% \hfill (5/5) & 100\% \hfill (5/5) & 75 & 0.0\% (0/161) & 4.4 \\
    S2095 & 46\% \hfill (361/782) & 9\% \hfill (34/361) & 4\% \hfill (34/782) & 170 & 0.6\% (1/161) & 6.3 \\
    S2111 & 100\% \hfill (69/69) & 53\% \hfill (37/69) & 53\% \hfill (37/69) & 185 & 1.2\% (2/161) & 4.9 \\
    S2116 & 100\% \hfill (1/1) & 100\% \hfill (1/1) & 100\% \hfill (1/1) & 5 & 0.0\% (0/161) & 4.5 \\
    S2142 & 99\% \hfill (315/316) & 95\% \hfill (300/315) & 94\% \hfill (300/316) & 4,500 & 0.0\% (0/161) & 4.6 \\
    S2184 & 97\% \hfill (431/440) & 85\% \hfill (368/431) & 83\% \hfill (368/440) & 1,840 & 0.6\% (1/161) & 4.5 \\
    S2225 & 13\% \hfill (3/22) & 100\% \hfill (3/3) & 13\% \hfill (3/22) & 15 & 0.6\% (1/161) & 4.5 \\
    S2272 & 97\% \hfill (40/41) & 85\% \hfill (34/40) & 82\% \hfill (34/41) & 170 & 0.0\% (0/161) & 4.5 \\
    S4973 & 98\% \hfill (80/81) & 85\% \hfill (68/80) & 83\% \hfill (68/81) & 340 & 1.2\% (2/161) & 4.4 \\
    \midrule
    ALL & 74\% \hfill (1,307/1,759) & 65\% \hfill (852/1,307) & 48\% \hfill (852/1,759) & 7,340 & 0.4\% (7/1,610) & -- \\
	\bottomrule
\end{tabular}
\begin{tablenotes}[flushleft]
    \item[a] Columns ``TV'', ``DV'', and ``FV'' indicate the numbers of \target violations, all detected violations, and fixed violations for each rule. Column ``TDR'' shows the percentage of detected violations that are \target. Columns ``FTR'' and ``FDR'' give the ratios of fixed \target and all detected violations (see formulas \eqref{eq:starget_fixed} and \eqref{eq:detected_fixed}). Column ``TRT'' indicates the estimated total remediation time for fixed violations. Column ``Failing\_Repos'' shows on how many projects there are tests failing because of \sorald. Column ``MT'' notes the median time spent by \sorald to fix violations of that rule in a single project.
\end{tablenotes}
\end{threeparttable}
\end{table*}

\autoref{tab:ex1_res} presents the results of our first experiment, which is aimed at assessing the applicability of \sorald. In this table, ``SQID'' is the identifier of \sonar rules. ``TV'' and ``DV'' represent the number of \target violations and all detected violations, respectively, as defined in \autoref{sec:protocol1}. The ``TDR'' column shows what percentage of detected violations are \target (see \eqref{eq:target_to_detected}). Also, ``FV'' shows the number of \emph{fixed violations} for each rule (see \autoref{sec:protocol1}). Based on these numbers, we compute ``FTR'' and ``FDR'', the ratio of fixed \target and all detected violations per  \eqref{eq:starget_fixed} and \eqref{eq:detected_fixed}. Using the number of fixed violations, we also calculate the estimated total remediation time for fixed violations, which is represented by ``TRT'' in the table. The ``Failing\_Repos'' column shows on how many projects (out of \topreposcnt) there are tests failing because of \sorald. Finally, for each rule, ``MT'' notes the median time (in seconds) spent by \sorald to fix violations of that rule in a single project.

\revision{Consider rule \textit{S1860} as an example. There are five detected violations (DV) of this rule over all \topreposcnt projects before running \sorald. All five are \target violations (TV) because they meet the \sorald assumptions. This means that $100\%$ of the detected violations of this rule can potentially be fixed by \sorald (TDR). Indeed, \sorald fixes all five \target violations, which means the \textit{FTR} and \textit{FDR} ratios for \textit{S2111} are both $100\%$. The estimated manual remediation time for those automatically fixed violations (\textit{TRT}), is 75 minutes for this rule. No project has a failing test after fixing the violations of \textit{S2111} with \sorald. Finally, the last column indicates that the median time that \sorald spends to fix violations of \textit{S2111} is $4.4$ seconds.}

\revision{Overall, 1,759 violations of all considered rules are detected in \toprepos, and 1,307 (74\%) of these detected violations are \target ones. This means most of the violations of the considered rules are appropriate to be fixed by our metaprogramming approach (see \autoref{sec:violation_fixing} for explanation).}

\revision{In total, 852 violations are fixed by \sorald, which represents $65\%$ of \target violations and $48\%$ of all detected violations. This percentage indicates \sorald's effectiveness at fixing almost two thirds of the \target violations. Moreover, the total remediation time (\textit{FRT}) of 7,340 minutes -- 122 hours of highly paid engineering hours -- suggests that practitioners can save a significant amount of time by using \sorald to repair \sonar violations, instead of doing it manually. We notice that for nine out of 10 considered rules, \sorald fixes a majority of \target violations. }

\revision{Having a closer look at the results, we see that \sorald's effectiveness significantly drops on violations of rule \textit{S2095} (used resources should be closed). Per our manual analysis, there are two facts that make automatically fixing violations of \textit{S2095} hard. First, it is not always straightforward to detect where a used resource should be closed. Second, there are instances where multiple resources are opened and not closed in a single block of code, which makes enclosing these resources in a try-with-resource complicated. Handling these two types of cases would require more fine tuning of our metaprogramming template for rule \textit{S2095}.}

\revision{Next, we investigate whether \sorald causes test regressions.
For all projects with at least one \sorald patch, we execute the whole test suite. Only seven of these (or 0.4\%) contain at least one test that fails after being patched by \sorald. These seven test failures occur in seven distinct projects. This means that in 99.6\% (1,603/1,610) of the cases, \sorald does not break the existing behavior of the program under repair to the extent it is specified by the tests. We manually analyze the seven  test failure and divide them into three groups as follows:}

\revision{First, in four of the seven cases where the test suite fails, the rule violation is intentional and fixing it breaks the expected behavior of the program. For example, in the ``kungfoo/geohash-java'' project the following statement exists: \texttt{``return 180d/Math.pow(2,bits/2);''}. Here developers intentionally divide \texttt{bits} by an integer ``2'' to get the floor of this division. \sorald breaks the expected behavior by changing this division to \texttt{bits/2D}, as the new expression no longer returns the floor of this division. This shows the importance of running tests after applying \sorald patches, as it can reveal some potentially unwanted fixes.}

\revision{Second, one of the test failures occurs because \sorald tries to fix a violation of \textit{S2095}, but the program semantics incorrectly changes after the fix. This happens because \sorald cannot correctly enclose the unclosed resources inside a try-with-resource block. Consequently, the piece of code that uses unclosed resources is eliminated and the functionality of the program is broken. Once again, this suggests the complexity of violations of rule \textit{S2095}}

\revision{Third, there are two other \sorald patches causing test failures that reveal undetected bugs. In these two cases, \sorald fixes a violation of \texttt{S2111} that is an actual bug in the program. Then, the behavioral change caused by the fix triggers a test failure. This means the tests are wrong and enforce the incorrect behavior of the buggy program before the \sorald fixes.}

\revision{Overall, there are only 7 cases out of 1,610 patches where \sorald impacts the test status of the project, which is arguably low. This suggests that \sorald can be trusted by developers to fix \sonar violations in their programs, as it usually does not break the tests.}

\revision{The total number of violations of all 97 fixable and partially fixable bug rules in our dataset is 4,928. Given that the number of all detected violations of our 10 considered rules is 1,759, we conclude that 35\% (1,759/4,928) of all violations of the fixable and partially fixable rules are covered by our considered rules. This suggests the relative representativeness of our selected rules.}

\revision{Finally, we discuss the performance of \sorald. As presented in \autoref{tab:ex1_res}, the median time spent by \sorald to fix violations of a considered rule on a project is $6.3$ seconds (this does not take into account the test suite execution).
Per our experience, the time required to run the tests and review the patch dominates the \sorald execution time.
Hence, \sorald itself is not a technical blocker compared to the environment and process considerations regarding automated fixing of static analysis warnings.}

\begin{mdframed}\noindent
    \textbf{Answer to RQ1: \rqone} \\
    \revision{According to our large scale evaluation, 65\% (852/1,307) of the \target violations over \topreposcnt projects are fixed by \sorald. This shows that \sorald is an effective method to repair violations of the considered rules. Also, \sorald is fast: it needs a median time of $6.3$ seconds to fix violations of a considered rule on a single project. This shows that \sorald can scale to the typical thousands of static analysis violations plaguing large scale software.}
\end{mdframed}


\subsection{Integration into Development Workflow (RQ2)}
\label{sec:ex_rq2}

\revision{Over the considered period of commit history of all \topreposcnt projects in \toprepos, \soraldbot finds there are 6,888 commits that are made in 126 repositories from Dec 12, 2020 to Nov 27, 2021. It means that 78\% (126/161) of \toprepos projects have been active during the last 350 days before conducting this experiment. Overall, this provides enough data for our experiment about estimating the impact of \soraldbot on real development workflows.}

\revision{Among 126 projects that have commits in the time interval under study, 16\% (21/126) of the projects have new \target violations in their changed files. These violations are introduced in 46 commits. This shows that even in a limited period of 350 days, there is a significant chance that an active project introduces new violations of the 10 considered rules, which reaffirms the usefulness of a tool like \sorald.}

\begin{table}[t]
\centering
\caption{\revision{RQ2 experimental results per rule. \soraldbot generates 54 patches that fix 80 newly introduced violations over 350 days. The numbers in the last row represent the number of unique items in the corresponding column. For example, there are 21 unique projects with new violations of at least one of the considered rules.}}
\label{tab:rq2_ex}
\begin{threeparttable}[b]
\begin{tabular}{l | r r r}
\toprule
	\textbf{SQID} & \textbf{\#Projects} & \textbf{\#Patches} & \textbf{FV} \\
	\midrule
	S2095 & 8 & 12 & 13 \\
	S2111 & 1 & 1 & 1 \\
	S2142 & 13 & 26 & 47 \\
	S2184 & 11 & 15 & 19 \\ 
	\midrule
	ALL & 21 & 54 & 80 \\ \hline
	\bottomrule
\end{tabular}
\end{threeparttable}
\end{table}

\revision{Among the 10 considered rules, four rules are violated in the studied interval: \textit{S2095}, \textit{S2111}, \textit{S2142}, and \textit{S2184}.} 
\autoref{tab:rq2_ex} presents the results for this experiment per rule. ``\#Projects'' is the number of projects that introduce new violations in the studied time interval. ``\#Patches'' represents the number of generated patches by \sorald. Note that for each rule that is violated in a commit, \soraldbot generates a single patch to fix all violations. The number of patches is the sum of the number of patches per rule. Finally, ``FV'' represents the number of fixed violations of the corresponding rule. \revision{For example, there are eight projects and 12 commits that are introducing new violations of rule \textit{S2095} in the studied period. In total, \soraldbot detects and fixes 13 violations of \textit{S2095} in the changed files of these 12 commits.}

\revision{As shown in \autoref{tab:rq2_ex}, \soraldbot generates 54 patches for 21 projects. Considering that these patches correspond to commits that are made over 350 days, we conclude that \soraldbot would not overwhelm the maintainers with too many patches. 41 patches fix exactly one violation and 13 patches fix between two and five violations. This shows the fixes for our 10 considered rules are usually short, which is in line with best practices for code review recommended by previous studies \cite{Gousios2014}.}

\begin{mdframed}\noindent
    \textbf{Answer to RQ2: \rqtwo} \\
    \revision{Over the considered period of 350 days, \soraldbot has seen 126 active projects in \toprepos with at least one commit, and has witnessed the introduction of violations of four different \sonar rules. This shows that even in a limited period of 350 days, opportunities do exist for \sorald to improve code quality. Notably, \soraldbot generates 54 patches that fix 80 newly introduced violations. This shows that \soraldbot is able to repair static analysis violations as soon as they are introduced.}
\end{mdframed}


\subsection{Action Research with Pull Requests (RQ3)}

\revision{\sorald is able to create fixes that are ready to review.
In total, we submit 29 pull requests to 21 \github projects, as shown in \autoref{tab:prs}. The size of this experiment is in line with related work \cite{MarcilioFBP20,carvalho2020c}. This provides us with a valuable insight regarding how useful developers consider \sorald. In this section, we present a qualitative analysis of our submitted pull requests.
We divide the submitted PRs into three groups: accepted (17 PRs), declined (10 PRs), and pending (2 PRs) and analyze the maintainers' reactions to each of these groups in the following sections.}

\begin{table}[t]
\label{tab:prs}
\centering
\scriptsize
\caption{\revision{RQ3: Pull requests submitted to real open source projects and their status.}}
\begin{tabular}{@{}l l r@{}}
\toprule
	ID & Pull request & Discussion \\
	\midrule
	PR1 & \href{https://github.com/DanielYWoo/fast-object-pool/pull/26}{DanielYWoo/fast-object-pool/pull/26} & \multirow{15}{*}{\makecell{Merged, \\ see Sec \ref{sec:rq3_merged}}} \\
	PR2 & \href{https://github.com/LinShunKang/MyPerf4J/pull/70}{LinShunKang/MyPerf4J/pull/70} &  \\
	PR3 & \href{https://github.com/LinShunKang/MyPerf4J/pull/71}{LinShunKang/MyPerf4J/pull/71} &  \\
	PR4 & \href{https://github.com/kafka-ops/kafka-topology-builder/pull/207}{kafka-ops/kafka-topology-builder/pull/207}	&  \\
	PR5 & \href{https://github.com/locationtech/jts/pull/811}{locationtech/jts/pull/811} &  \\
	PR6 & \href{https://github.com/pinterest/singer/pull/148}{pinterest/singer/pull/148} &  \\
	PR7 & \href{https://github.com/rawls238/Scientist4J/pull/42}{rawls238/Scientist4J/pull/42} &  \\
	PR8 & \href{https://github.com/vert-x3/vertx-jdbc-client/pull/258}{vert-x3/vertx-jdbc-client/pull/258} &  \\
	PR9 & \href{https://github.com/networknt/json-schema-validator/pull/488}{networknt/json-schema-validator/pull/488} &  \\
	PR10 & \href{https://github.com/jMetal/jMetal/pull/442}{jMetal/jMetal/pull/442} &  \\
	PR11 & \href{https://github.com/jMetal/jMetal/pull/443}{jMetal/jMetal/pull/443} &  \\
	PR12 & \href{https://github.com/lolo101/MsgViewer/pull/65}{lolo101/MsgViewer/pull/65} &  \\
	PR13 & \href{https://github.com/lolo101/MsgViewer/pull/66}{lolo101/MsgViewer/pull/66} &  \\
	PR14 & \href{https://github.com/LianjiaTech/retrofit-spring-boot-starter/pull/84}{LianjiaTech/retrofit-spring-boot-starter/pull/84} &  \\
	PR15 & \href{https://github.com/LianjiaTech/retrofit-spring-boot-starter/pull/85}{LianjiaTech/retrofit-spring-boot-starter/pull/85} &  \\
	PR16 & \href{https://github.com/carml/carml/pull/110}{carml/carml/pull/110} &  \\
	PR17 & \href{https://github.com/wmixvideo/nfe/pull/776}{wmixvideo/nfe/pull/776} &  \\
	\midrule
	PR18 & \href{https://github.com/AddstarMC/Prism-Bukkit/pull/247}{AddstarMC/Prism-Bukkit/pull/247} & \multirow{10}{*}{\makecell{Declined, \\ see Sec \ref{sec:rq3_rejected}}} \\ 
	PR19 & \href{https://github.com/AddstarMC/Prism-Bukkit/pull/248}{AddstarMC/Prism-Bukkit/pull/248} &  \\
	PR20 & \href{https://github.com/Athou/commafeed/pull/957}{Athou/commafeed/pull/957} &  \\
	PR21 & \href{https://github.com/CloudburstMC/Nukkit/pull/1764}{CloudburstMC/Nukkit/pull/1764} &  \\
	PR22 & \href{https://github.com/kafka-ops/kafka-topology-builder/pull/206}{kafka-ops/kafka-topology-builder/pull/206} &  \\
	PR23 & \href{https://github.com/techa03/goodsKill/pull/79}{techa03/goodsKill/pull/79} &  \\
	PR24 & \href{https://github.com/techa03/goodsKill/pull/80}{techa03/goodsKill/pull/80} &  \\
	PR25 & \href{https://github.com/alibaba/arthas/pull/2037}{alibaba/arthas/pull/2037} &  \\
	PR26 & \href{https://github.com/failsafe-lib/failsafe/pull/314}{failsafe-lib/failsafe/pull/314} &  \\
	PR27 & \href{https://github.com/failsafe-lib/failsafe/pull/315}{failsafe-lib/failsafe/pull/315} &  \\
	\midrule
	PR28 & \href{https://github.com/xerial/sqlite-jdbc/pull/696}{xerial/sqlite-jdbc/pull/696} & \multirow{4}{*}{\makecell{Pending, \\ see Sec \ref{sec:rq3_pending}}} \\
	PR29 & \href{https://github.com/cmu-phil/tetrad/pull/1355}{cmu-phil/tetrad/pull/1355} &  \\
	\bottomrule
\end{tabular}
\end{table}

\subsubsection{Case: Accepted}
\label{sec:rq3_merged}

\begin{lstlisting}[float=tb, style=diff, caption={Diff from an accepted pull request (\textit{PR1}) that fixes violations of \textit{S2142} in a recently changed Java file of project ``fast-object-pool''.}, captionpos=b, label=lst:merged_pr1]
    Thread.sleep(1000L * 2);
} catch (InterruptedException e) {
    e.printStackTrace();
%\GHilight%+   Thread.currentThread().interrupt();
}
System.out.println();
\end{lstlisting}

\revision{Seventeen of the 29 submitted pull requests are accepted. Nine of the accepted pull requests fix violations of \textit{S2142} (wrong thread API usage, see \autoref{tab:rules}). Four other PRs fix violations of \textit{S2184}  (math operand type mismatch). One PR fixes violations of \textit{S2111} (wrong instantiation of \texttt{BigDecimal}). Finally, three PRs fixes violations of \textit{S2095}  (unclosed resources).}

\autoref{lst:merged_pr1} shows the  \textit{PR1} patch, which fixes a violation of \textit{S2142} in ``fast-obj-pool''. To fix the violation, this PR adds one line which interrupts the current thread, per the \sonar's official guideline. After the PR was accepted, we engaged in a conversation with one of the maintainers of ``fast-obj-pool'' to gain more insights. In our conversation, the maintainer noted that the issue fixed by \sorald is ``an obvious issue'' confirming we do not need to provide a complicated description for our PR. \revision{This comment supports our take that the simplicity of \sorald's patches plays a role in the acceptability of submitted PRs.}

\textit{PR4} fixes violations of rule \textit{S2095}, which is tagged as a ``blocker'' rule in \sonar documents: \textit{S2095} tags unclosed resources, which can result in resource leak. A blocker rule has the highest severity level among \sonar rules. \revision{This PR  has been accepted.}

Finally, \textit{PR3} fixes violations of \textit{S2184} in ``MyPerf4J''. This PR was pending for a significant time, until we noticed that the PR violated two checkstyle checks on the project. Two weeks after submitting the original PR, we manually fixed the style issues to see if we receive a decision from the maintainers. Two days later, the pull request was accepted. The story of this pull request confirms that for some projects, there are non-behavioral requirements (such as formatting), enforced in continuous integration, that are mandatory.

\subsubsection{Case: Declined}
\label{sec:rq3_rejected}

\begin{lstlisting}[float=tb, style=diff, caption={A \sorald fix for a violation of \textit{S2184} in the ``Prism-Bukkit'' project. This patch is deemed unnecessary by the developer because it applies to fixed literal values making the violation actually impossible (integer overflow).}, captionpos=b, label=lst:rejected_pr1]
%\RHilight+% long period = 24 * 60 * 60;
%\GHilight+% long period = 24L * 60 * 60;
\end{lstlisting}

\revision{Three of the declined pull requests fix violations of \textit{S2184}  (math operand type mismatch). \autoref{lst:rejected_pr1} shows \textit{PR18} submitted to ``Prism-Bukkit''.} In this violation, three \texttt{int} literals are multiplied and the result is saved in a \texttt{long} variable. According to \sonar's official documentation, \sorald proposes to cast one of the \texttt{int} numbers to \texttt{long} to avoid an overflow. This patch fixes the \sonar violation, but it is unnecessary because $24*60*60$ is 86,400, which is well below the upper range of Java's \texttt{int} ($2^{31}-1$). That is the point noted by the maintainer of this project on the pull request. \revision{The same problem also applies to \textit{PR21} submitted to ``Nukkit'' and \textit{P24} submitted to ``goodsKill''.} 
Overall, we conclude that \sorald's fixes for rule \textit{S2184} are not undoubtedly valuable, because violations of this \sonar rule are often false positives. To overcome this issue, future work can be made to filter out violations before repairing them.

There are also \textit{PR19}, \textit{PR22}, and \textit{PR25} that fix their targeted violations in a way that is not acceptable by developers. First, \textit{PR19} fixes a \textit{S2095} violation in ``Prism-Bukkit'' and it is not accepted. A part of the maintainer's comment on this fix is this: ``yes you [are] right but ... that rule is meant to be interpreted.''. This shows that the maintainer accepts that the violation is correctly detected, but he is not satisfied with the fix because it does not provide the correct behavior. \revision{The same thing happens with \textit{PR25}. In \textit{PR25}, the developer declines our pull request to fix a violation of \textit{S2095} but mentions a new PR that fixes, in a different way, the same violation and that was created by himself. Rule \textit{S2095} is related to unclosed resources. As explained in \autoref{sec:rq1_results}, fixing its violations is non-trivial and needs advanced data flow analysis. Moreover, the case of \textit{PR25} shows that the developer might come up with a different solution for the same violation. Yet, our pull request to fix violations of the very same rule in ``kafka-topology-builder'' (\textit{PR4}) has been accepted and merged by the maintainers of that project, which shows that considering this rule in \sorald brings value, but in a context-dependent way.}

\revision{There is also \textit{PR22} that is declined because the maintainers think the violation should be fixed differently.} This PR fixes violations of \textit{2142} in the ``kafka-topology-builder'' project. In reaction to the PR, a main contributor of this project says: ``I don't think this is the proper fix for this issue'' and makes alternative suggestions. Then, the maintainer proceeds saying that: ``The risk is that because it is an automated patch, many developers will think that this [is] the right/only way and just approve it as long as the tests are green''. 
This deep qualitative feedback shows that:
1) \soraldbot PRs should be manually reviewed by developers before getting accepted.
2) Some developers are skeptical about the whole idea of fixing \sonar violations automatically.
\revision{Yet, the same developer accepted \textit{PR4} which repairs a different rule, showing that fine-tuned and well targeted fixes can even convince developers skeptical of some aspects of \sorald to use it.}

\revision{Finally, \textit{PR20} fixes violations of \textit{S2142} in the ``commafeed'' project. The maintainer of the project declined the pull requests without any comments, and they did not respond to our question as of Dec 17, 2021. Looking at the commit history of this project, we see that this project is not active and does not welcome external contributions.}

\subsubsection{Case: Pending}
\label{sec:rq3_pending}

\revision{There are two pull requests that are still pending maintainer decisions as of Dec 17, 2021. These PRs are submitted to ``sqlite-jdbc'', and ``cmu-phil''. Looking at the accepted PRs on ``cmu-phil'' during the past year, we notice that all of them are created either by a well known software bot, such as ``dependabot'' or by the maintainers of the project. Therefore, we conclude that this project is not an appropriate subject for this experiment.} 

\subsubsection{Lessons Learned} 
By analyzing all the reactions from developers of real-world open source software, we are able to make the following observations.
\revision{\begin{itemize}
    \item \emph{Simple is better.} The maintainer reaction on \textit{PR1} suggest that fixing only one rule per PR is better, and doing it with a short patch allows for fast understanding, code reviewing and merging.
    \item \emph{Severity is important.} The PRs that fix severe violations are more welcome by developers. Our future engineering work on \sorald will mainly focus on rules that are known to be more severe to make sure the \sorald patches are welcomed by developers (see \textit{PR4} and \textit{PR27}).
    \item \emph{Context matters.} Violations of the same rule may require different fixing strategies. Hence, having more than one metaprogramming templates for each rule and picking one of them based on the context of the target program is a promising direction of future work (see \textit{PR19} and \textit{PR22}).
    \item \emph{Functional behavior is not everything.} Many projects have non-behavioral requirements for PRs to be accepted, some of them ensured in continuous integration. One example is code style requirements. To maximize acceptance, \sorald suggestions must take all non-behavioral aspects into account (see \textit{PR3}, \textit{PR9}, and \textit{PR16}).
\end{itemize}}

\begin{mdframed}\noindent
    \textbf{Answer to RQ3: \rqthree} \\
    By doing action research with real pull requests, we are able to obtain deep qualitative insights on the relevance of \sorald's automated patches. Our analysis of developer reactions to \sorald's pull requests reveals that focusing on severe violations is of utmost importance.
    \revision{We also find that considering non-functional requirements of target projects, like code style, is an important aspect of contributions by software bots.}
\end{mdframed}

\section{Discussion}

\subsection{Comparison with \spongebug}
\label{sec:spongebugs_comparison}
Among the related work, Marcilio et al.'s work \cite{MarcilioFBP20} is particularly close to our study. They introduce \spongebug which is a program transformation technique that fixes violations of 11 \sonar rules. \spongebug is similar to \sorald in that they both apply predefined transformation templates on a program to repair \sonar warnings. Note that \spongebug does not offer a repair bot to integrate their tool into development workflows.
In addition to this, \sorald and \spongebug have the following major differences:

\textit{Target rules:} \spongebug considers 11 most common rules, mostly of importance ``code smell'' (9/11 rules). On the other hand, we only focus on rules of importance ``bug''. Our paper is the first comprehensive study of automatic repair for \sonar ``bug'' rules, beyond code smells.

\revision{\textit{Reliability and Extensibility:} In \spongebug, the detection of rule violations is implemented from scratch. In \sorald, we do not reimplement it, we reuse it the implementation provided by a static analyzer. This brings two advantages for \sorald. First, the reliability of the rule violation detection is high, because it is done by professional developers and has been battle-tested over thousands of projects. Second, it makes \sorald easier to extend for repairing additional rules. Once a static analyzer is plugged into \sorald (like \sonar currently is), the support for a given rule only requires the repair part, and not the detection while adding a new rule to \spongebug requires the implementation of both the detection and the repair parts.}

\textit{Effectiveness:} To compare the effectiveness of \sorald with \spongebug, we run \sorald on the dataset that Marcilio et al. utilize to evaluate \spongebug applicability. This dataset includes 12 well established open source projects, such as Eclipse Platform UI. We take the intersection of \spongebug and \sorald's rules, consisting of two bug related rules, namely \textit{S4973} and \textit{S2111}.

\begin{table}[t]
\centering
\caption{\sorald's and \spongebug's effectiveness on the intersection of supported rules (two rules in common).}
\label{tab:comparison_spongebug}
\begin{tabular}{l | r r r}
\toprule
	\textbf{Tool} & \textbf{S4973} & \textbf{S2111} & \textbf{Total} \\
	\midrule
	\sorald & 93\% \hfill (92/99) & 100\% \hfill (14/14) & 94\% \hfill (106/113) \\
	\spongebug & 53\% \hfill (52/99) & 71\% \hfill (10/14) & 54\% \hfill (62/113) \\
	\bottomrule
\end{tabular}
\end{table}

\autoref{tab:comparison_spongebug} shows the results of comparison between \sorald and \spongebug on the two rules commonly implemented. Each cell of this table has three numbers with the following format: P\%~(X/Y). Y is the number of violations of that rule in the dataset, X is the number of violations fixed by the corresponding tool, and P is the percentage of fixed violations.

In total, this dataset has 99 violations of \textit{S4973} and 14 violations of \textit{S2111}. For both rules, \sorald has a higher fix rate. For \textit{S4973}, it is 93\% vs 53\% and for \textit{S2111} it is 100\% vs 71\%. In total, \sorald fixes 94\% (106/113) of the violations, where \spongebug fixes 54\% (62/113). These results suggest that \sorald is more effective than \spongebug at resolving \sonar bug related violations.

\revision{To better understand why \sorald outperforms \spongebug, we compare the transformation templates and violation fixes of these two rules (\textit{S4973} and \textit{S2111}) as follows:}
\revision{We manually compare the implementation of transformation templates that \sorald and \spongebug employ. We conclude that both tools use the same template concept. To fix violations of \textit{S4973}, any comparison between two strings or boxed types using the \texttt{==} operation should be replaced with \texttt{equals()}. For violations of rule \textit{S2111}, both \sorald and \spongebug change instantiations of \texttt{BigDecimal} objects that are implemented as \texttt{new BigDecimal(X)} with an instantiation implemented as \texttt{BigDecimal.valueOf(X)}. This shows that for both rules, \sorald and \spongebug use the same transformation templates.}

\revision{Since the transformation templates used by the two tools are conceptually the same, the effectiveness difference between the tools necessarily comes from the implementation details. To investigate the difference, we also compare the particular violations that are fixed by both tools with violations that are fixed by \sorald but not by \spongebug. Our manual analysis shows the violations that are not fixed by \spongebug appear in more complex code expressions. For example, \spongebug fixes instantiations of a \texttt{BigDecimal} object that look like \texttt{new BigDecimal(doubleValue)}. However, when the same instantiation is followed by a method invocation, like \texttt{new BigDecimal(doubleValue).toBigInteger()}, it is not fixed. Detecting why this happens requires an intensive debugging of \spongebug, which is out of scope of this study. Nonetheless, since we know the transformation templates used by \sorald and \spongebug are the same, we conclude that \spongebug likely misses some violations that can be fixed by \sorald. Recall that \sorald integrates \sonar violation detection into its own framework, while \spongebug implements the detection mechanism on its own. This gives \sorald a superior detection ability.}

\subsection{Threats to Validity}
\label{sec:threats}

\revision{\textit{Correctness assessment:} \sorald's fixes are designed to not break the intended functionality of target programs. After a \sorald patch, we always run all the tests of the considered project.
However, it is well known that passing all the tests does not necessarily entail having a correct behavior \cite{ye2021automated}. To make sure that the behavior of a program is correct, as discussed in \autoref{sec:integration-workflow}, \sorald's patches are meant to be reviewed by developers who are familiar with the application domain and its requirements. 
In RQ1, we have 852 fixed violations with \sorald  and show that in 99.6\% of cases, no tests break after repair. It was arguably not feasible to ask the maintainers of the target programs to label all 852 fixes as correct/incorrect. Therefore, there is a theoretical risk that the 99.6\% figure of passing tests is an over-approximation of the ability of \sorald to not break expected behavior.}

\revision{\textit{Rule labeling:} As described in \autoref{sec:considred_rules}, we label all \sonar bug rules as fixable, partially fixable, and unfixable. As the labeling process includes a manual analysis based on the participants' expertise, it can be error prone in some cases. In particular, analyzers have to determine if the common expected behavior of developers can be implemented by applying metaprogramming templates. This is a fuzzy question and participants' answer to it heavily depends on their experience and understanding of software. To overcome this threat, two different authors with considerable experience in program repair conduct this analysis and resolve their conflicting labels through detailed discussions.}

\section{Related Work}
\label{sec:related}
There are many common static code analyzers that perform static analysis on programs to identify potential bugs, vulnerabilities, or code smells \cite{zampetti2017how}, such as \sonarqube, \textsc{SpotBugs} \cite{lavazza2020empirical}, \textsc{FindBugs}\cite{hovemeyer2004finding}, and \textsc{PMD} \cite{trautsch2020longitudinal}. 

In the following, we review studies on the usage of static analyzers and its effects. Next, we discuss two types of tools: tools that propose fix suggestions for static warnings and program repair bots that are integrated into development platforms via pull requests.


\subsection{Usage of Static Code Analysis}
Many researchers have studied the usage of static analysis and its effects. In one line of study, researchers investigate whether developers act on the warnings produced by static analyzers \cite{imtiaz2019developers,johnson2013don}. It is shown that a low ratio of static analyzer warnings are repaired by developers \cite{MarcilioBMCL019} and two main reasons are presented for this. First, the warnings do not correspond to actual program flaws (i.e., false positive warnings) \cite{ayewah2010google,MarcilioBMCL019}. Second, developers prefer to work on warnings that they know how to fix \cite{do2020software}, while static analyzers do not always present warnings properly to help developers in this regard \cite{johnson2013don}. These drawbacks have made it costly for developers to fix program failures detected by static analyzers \cite{sadowski2018lessons}. Consequently, developers are not likely to have strict policies on the static analyzer use in their projects \cite{beller2016analyzing}.

Due to the prevalence of false positive warnings, many developers look for ways to filter out warnings that do not have real impact on programs \cite{imtiaz2019challenges}. Lenarduzzi et al. \cite{LenarduzziS020} argue that developers cannot determine the actual impact of a \sonarqube warning only by looking at the severity level assigned by the tool. Because of this, several methods have been proposed to detect false positive warnings automatically \cite{muske2020techniques}. For example, Yang et al. \cite{yang2020understanding} introduce an incremental support vector machine mechanism to distinguish between warnings that represent serious problems in programs and unimportant warnings.

To address the second reason behind static analyzer warnings being ignored, researchers look for new methods to present warnings to developers. Using static analyzers in continuous integration (CI) platforms is one of such methods. In this regard, Zampetti et al. \cite{zampetti2017how} find that CI builds almost never fail due to a potential bug detection by a static analyzer. They argue one of the main reasons is that developers who use static analyzers do not configure them to break builds because they do not want to be forced to cope with the warnings. Vassallo et al. \cite{vassallo2018continuous} also report that developers do not perform continuous code quality checks over time, rather they do it at the end of development sprints. This finding confirms the usefulness of \sorald in helping developers handle static warnings as early as possible.

Baldassarre et al. \cite{BaldassarreLRS20} investigates the accuracy of \sonarqube prediction of the remediation time, the time required for fixing the detected technical dept items and conclude that \sonarqube usually overestimates the remediation time. This suggests that more research is required to have an accurate assessment of the time that can be saved by using static warning repair tools.


\subsection{Automatic Fix Suggestion for Static Warnings}

Bader et al. \cite{BaderSP019} introduce an approach, called \textsc{Getafix}, to fix bugs detected by two static analyzer tools, Infer \cite{CalcagnoDDGHLOP15} and Error Prone \cite{AftandilianSPK12}. First, they employ a hierarchical clustering algorithm to extract a hierarchy of fix patterns from general to specific ones from their training dataset. Next, they use these fix patterns to make fix suggestions. Finally, they rank the fix suggestions and recommend the best ones to the developer. \textsc{Getafix} is different from \sorald in that: 1) it fixes bugs detected by Infer and Error Prone which are not as widely used in industry as \sonarqube 2) it uses a data-driven approach as opposed to a metaprogramming one.

\textsc{Phoenix}~\cite{YoshidaBHNPK20} is another tool for resolving static warnings. \textsc{Phoenix} has three stages for fixing SpotBugs violations.
\begin{enumerate*}
\item Collecting violation fixing patches.
\item Learning repair strategies.
\item Suggesting repairs.
\end{enumerate*}
It takes advantage of \textsc{GumTree} \cite{FalleriMBMM14} to compute AST differences and \textsc{Eclipse JDT} to manipulate ASTs. While Phoenix is a learning-based tool, \sorald is rule based.

Logozzo and Ball \cite{logozzo2012modular} propose an approach to repair programs that violate some specifications in .NET programs, such as assertions, preconditions, and runtime guards. Their approach exploits the semantic knowledge of a program, which is given by abstract states, to produce repaired versions. This approach is implemented as an extension for \textsc{cccheck} \cite{fahndrich2010static}. \sorald generates the fix suggestions by transforming the program AST according to predefined metaprogramming templates, while \textsc{cccheck} suggests the repairs based on an analysis of possible abstract states of the program.

Liu et al. \cite{liu2019avatar} introduce \textsc{AVATAR}. \textsc{AVATAR} mutates programs with test failures to generate a patch that passes all the tests. To do this, \textsc{AVATAR} applies fix patterns that are mined from a large dataset \cite{liu2018mining}. To extract the patterns, they first find code changes that remove violations of \textsc{FindBugs}. Then, they use \textsc{GumTree} \cite{FalleriMBMM14} to extract AST edit scripts in these changes. Next, they use a deep learning technique to cluster the edit scripts based on their similarity, and finally, manually design specific fix patterns based on edit scripts in each cluster. Each of these fix patterns repair violations of specific rules of \textsc{FindBugs}. Liu et al. \cite{liu2019avatar} show the effectiveness of \textsc{AVATAR} by testing it on the Defects4J dataset. The main difference between \sorald and \textsc{AVATAR} is that they have different goals: \sorald intends to fix violations of a static analyzer, while \textsc{AVATAR} applies violation fixing patterns to generate a test passing patch.

\revision{\textsc{TFix} \cite{berabi2021tfix} takes advantage of a learning-based approach to fix static warnings. It formulates fixing as a text-to-text task that uses neural networks to translate a rule violating program to a violation free version. \textsc{TFix} specifically fixes violations of \textsc{ESLint} rules in JavaScript programs. It is different from \sorald in the sense that it targets warnings of a different static analysis tool and it uses a learning-based approach, rather than a template-based one, to fix warnings.}
\spongebug \cite{MarcilioBMCL019} is the closest related work, and we have stated in \autoref{sec:spongebugs_comparison} the major differences with \sorald.


\subsection{Program Repair Bots}
As program repair technology becomes more advanced, there is a great motivation for researchers to find a way to integrate such technologies into common development workflows. For this purpose, van Tonder and Le Goues \cite{van2019towards} envision repair bots that orchestrate automated refactoring and bug fixing. They discuss six syntax and semantic related principles that researchers should take into account for engineering repair bots.

Urli et al. \cite{urli2018design,monperrus2019repairnator} introduce \textsc{Repairnator}, as a program repair bot. \textsc{Repairnator} constantly monitors \textsc{Travis-CI} builds on \github repositories. When a build fails, \textsc{Repairnator} reproduces the failure locally, uses repair tools to patch the program, finds a plausible patch that passes all the tests, and submits the passing patch as a pull request to maintainers of the failing project. \textsc{Repairnator} is designed to be extensible: one can add new repair tools to generate patches, or define new conditions for a plausible patch. 

\textsc{SapFix} is a program repair bot introduced by Marginean et al. \cite{marginean2019sapfix} at Facebook. \textsc{SapFix} is integrated into the \textsc{Phabricator} continuous integration platform. Once a new diff is submitted on \textsc{Phabricator}, \textsc{SapFix} employs three strategies: mutation-based, template-based, and revert-based (which just reverts to the old version) changes to generate a set of patches that pass all tests. Next, the patches that pass all the tests are prioritized, and the top one is reported to developers as a fix. The main difference between \soraldbot and \textsc{SapFix} is that \soraldbot tries to fix \sonarqube violations, while \textsc{SapFix} fixes programs with null pointer exceptions. 

The repair bot that is the most similar to \soraldbot is \textsc{C-3PR} \cite{carvalho2020c}. \textsc{C-3PR} monitors pushes on Git repositories and looks for changed files with static warnings. Next, it applies appropriate transformations to fix the warnings. If a warning is fixed by transformations, \textsc{C-3PR} submits a pull request including the fix. In contrast with \soraldbot, \textsc{C-3PR} does not have its own code transformation or violation detection technique. Instead, it uses that of \textsc{ESLint}, \textsc{TSLint}, and \textsc{WalkMod}. The major advantage of \soraldbot is to target the industry standard \sonar platform.

\revision{Serban et al. \cite{serban2021saw} introduce \textsc{SAW-BOT}, which fixes five code smell \sonar rules using JavaParser. The authors compare three ways of suggesting fixes. First, the legacy mode in which all violations in the whole project are fixed and the fixes are suggested in a pull request. Second, the pull request mode in which only violations in a recently created pull request are fixed and then suggested in a separate pull request. Third, only violations in a pull request are fixed and suggested as part of a code review. The authors perform a small study and conclude that developers prefer the third option because they can focus on reviewing the soundness of the fix rather than on understanding why the fix is there. Moreover, the third option, which the authors call ``Github Suggestions'' mode, gives developers more control because in this mode suggestions are made directly into the pull request. \textsc{SAW-BOT} is different from \sorald as the former focuses on codesmell rules, while \sorald fixes violations of bug rules. Nonetheless, \soraldbot can take advantage of Serban et al.'s conclusion and submit its fixes as a part of a code review.}

There are also a number of research works about refactoring bots. Wyrich and Bogner \cite{wyrich2019towards} propose a prototype tool that suggests fixes for a few types of \sonarqube code smell violations. This tool is meant to be triggered manually by an API call. Alizadeh et al. \cite{alizadeh2019refbot} introduce \textsc{RefBot}, \textsc{RefBot} monitors pull requests on a specified \github repository. For each PR, it computes different code quality metrics on changed files to see if there is a refactoring opportunity. If it finds a refactoring opportunity, \textsc{RefBot} adopts a set of predefined refactoring operations to improve the code quality. For successful refactorings, \textsc{RefBot} submits a pull requests to suggest it to the developers.

Overall, the main novelty of \soraldbot is that it is the first repair bot for static warnings which has its own repair engine.

\section{Conclusion}
\label{sec:conclusion}
In this paper, we introduce \sorald, a novel system to automatically fix violations of \sonarqube rules. \sorald considers 10 \sonarqube rules that correspond to potential bugs. To fix violations, \sorald has a predefined metaprogramming template that it applies to the target program. This template transforms a rule violating program to a violation-free one based on transformation of the abstract syntax tree.

\revision{We evaluate \sorald's applicability on \topreposcnt popular open source projects containing 1,307 \target violations of the considered rules. The results show that 65\% (852/1,307) of these violations are fixed by \sorald, which is a promising result.} Moreover, we propose a method to integrate \sorald into well known development workflow with pull requests. Our assessment of this integration method shows that \sorald is able to make an appropriate amount of useful contributions by monitoring recent commits, detecting violation introducing changes, and proposing fixes for introduced changes in the form of pull requests.

In the future, researchers can look for effective methods to automatically predict \sorald patches that are not likely to be accepted by developers for various reasons. One option is to build a machine learning model based on historical data related to repaired and unrepaired \sonar violations and their fixes to determine if a \sorald patch is likely to be accepted by developers. Such models enable us to filter out less useful \sorald suggestions to avoid wasting development resources by asking developers for unnecessary reviews.

\revision{
\section{Acknowledgments}
We acknowledge the important contribution of Aman Sharma, who is a Research Engineer at KTH Royal Institute of Technology, to the engineering of Sorald in the late stage of the project. This work was partially supported by the Wallenberg Artificial Intelligence, Autonomous Systems and Software Program (WASP) funded by Knut and Alice Wallenberg Foundation, and by the Swedish Foundation for Strategic Research (SSF). Some experiments were performed on resources provided by the Swedish National Infrastructure for Computing.
}

\balance
\bibliographystyle{IEEEtran}
\bibliography{references}

\end{document}